\documentclass{article}
\usepackage[utf8]{inputenc}
\usepackage{amsmath}
\usepackage{algorithm}
\usepackage{algorithmicx}
\usepackage{amsfonts}
\usepackage[a4paper]{geometry}
\usepackage{enumitem}
\usepackage{color}
\usepackage{hyperref, cite}
\usepackage{cleveref}
\usepackage{listings}
\usepackage{titlesec}
\usepackage{graphicx} 
\usepackage{placeins}
\usepackage{float}     
\usepackage{caption, authblk}  
\usepackage{subcaption}
\usepackage{algpseudocode}

\title{Microstructure and Manipulation: Quantifying Pump-and-Dump Dynamics in Cryptocurrency Markets}
\author{Mahya Karbalaii}
\affil{\small{LUISS Data Lab, LUISS Guido Carli, Viale Pola, 12, 00198 Roma, Italy\\ mkarbalaii@luiss.it}}
\date{April 2025}

\begin{document}

\maketitle

\section*{Abstract}
Building on our prior threshold‐based analysis of six months of Poloniex trading data, we have extended both the temporal span and granularity of our study by incorporating minute‑level OHLCV records for 1,021 tokens around each confirmed pump‑and‑dump event. First, we algorithmically identify the accumulation phase—marking the initial and final insider volume spikes—and observe that 70\% of pre‑event volume transacts within one hour of the pump announcement. Second, we compute conservative lower bounds on insider profits under both a single‑point liquidation at 70\% of peak and a tranche‑based strategy (selling 20\% at 50\%, 30\% at 60\%, and 50\% at 80\% of peak), yielding median returns above 100\% and upper‐quartile returns exceeding 2000\%. Third, by unfolding the full pump structure and integrating social‑media verification (e.g., Telegram announcements), we confirm numerous additional events that eluded our initial model. We also categorize schemes into “pre‑accumulation” versus “on‑the‑spot” archetypes—insights that sharpen detection algorithms, inform risk assessments, and underpin actionable strategies for real‑time market‑integrity enforcement.\\

\textbf{Keywords:} Cryptocurrency, Market Manipulation, Pump and Dump, Disinformation, Financial Scam

\section{Introduction}
Cryptocurrency markets, by virtue of their 24/7 operation, low liquidity, and jurisdictional fragmentation, are particularly susceptible to coordinated market manipulations \cite{crypto_crime}. Among these, pump-and-dump (P\&D) schemes stand out for their rapid price inflation, driven by organized insider purchases, and subsequent collapses that inflict heavy losses on other participants \cite{sec, fatf}. Unlike traditional equities, where regulatory oversight and circuit breakers curtail such practices, crypto exchanges often lack enforcement mechanisms, enabling malicious actors to exploit information asymmetries and technological affordances \cite{Karbalaii2025Crypto}.

Historically, pump-and-dump manipulation dates back to early 20th-century penny-stock frauds, and re-surged during the dot‑com bubble when unscrupulous brokers hyped illiquid internet shares \cite{to_the_moon, anatomy_pump}. In the crypto realm, the phenomenon has been supercharged by messaging platforms such as Telegram and Discord, which facilitate anonymous coordination of thousands of participants \cite{to_the_moon, doge2023pump}. Prior research \cite{Karbalaii2025PumpDetect} introduced an unsupervised, threshold-based model using hourly Poloniex data — combining exponentially weighted moving averages, volatility measures, and double-conditioned volume thresholds — to detect local price–volume anomalies. While effective in flagging coarse-grained events, that framework could not resolve the sub-hour insider strategies underpinning each pump.

In order to better explain the structure of pump and dump events and quantify the behavioral dynamic from the organizers, we provide an extension to the previous work \cite{Karbalaii2025PumpDetect}.
In the current article we address the following critical gaps:
\begin{enumerate}
  \item \textbf{Temporal Precision in Accumulation Phase:} We refine the analysis window to minute-level resolution, capturing the exact timing of insider volume spikes and observing that accumulation is often compressed into the last 24 hours before the pump announcement.
  \item \textbf{Insider Profit Quantification:} Leveraging first-trade and VWAP proxies for purchase cost, combined with conservative liquidation scenarios (70\% of peak or staged tranches at [50\%,60\%,80\%]), we derive lower-bound profit distributions that underscore the immense incentives driving P\&D operations.
\end{enumerate}
By integrating these elements and validating against social-media–confirmed events, we demonstrate improved detection accuracy and uncover previously undetected pump occurrences. The remainder of the paper is structured as follows: Section \ref{sec:data} details data sources and pre-processing; Section \ref{sec:accumulation} and \ref{sec:quantification} analyze accumulation-phase dynamics; Section \ref{sec:profit} introduces profit estimation methods and results; Section \ref{sec:conclusion} concludes with implications and future work.

\section{Data and Event Selection}
\label{sec:data}
We obtained our dataset from Poloniex’s public API \cite{poloniexapi}, comprising 1,101 trading pairs over the six‑month period from 15 August 2024 to 15 February 2025.  Applying the hourly threshold model \cite{Karbalaii2025PumpDetect}, we identified 1,021 tokens with at least one flagged event (hereafter “Target Date”).  For each token–event pair (grouped by \{\texttt{symbol}, \texttt{target\_date}\}), we extracted minute‑level OHLCV data from four days before to two days after each event, yielding approximately 8.2 million records.  All timestamps were converted to UTC.

Some tokens exhibited multiple flagged events and thus appear under more than one Target Date.  Our initial detection model employed hourly data to reduce market noise and successfully flag candidate pump‑and‑dump episodes.  Building on this, our objective was to “zoom in” around these detected episodes, leveraging higher‑resolution data, to uncover finer patterns that reliably confirm event occurrence.

Previous work suggests that pump‑and‑dump (P\&D) schemes unfold in three phases: accumulation, pump, and dump \cite{crypt_schemes, lee2008pump, crypto_pump_scheme}.  During the accumulation phase, organizers covertly accumulate the target token at low prices without attracting market attention.  However, several practical considerations constrain the duration of this phase:
\begin{itemize}
    \item \textbf{Speculative trading style:} P\&D organizers treat these operations as short‑term speculation, not long‑term investment, and hence prefer not to tie up capital for extended periods.
    \item \textbf{Risk of counter‑schemes:} Early accumulation increases the risk that other organizers might target the same token, potentially forcing the initial accumulators to buy at inflated prices or suffer losses.
\end{itemize}
On these grounds, we posit that the accumulation phase most likely occurs within four days preceding the Target Date.  Although the pump and dump themselves typically conclude within minutes, we conservatively include a two‑day post‑event window to capture any residual market effects.  

\section{Accumulation Phase}
\label{sec:accumulation}
To characterize and quantify the accumulation phase within our dataset, we performed targeted visual analyses on several subsets of 20 tokens. For each selected token, we plotted minute‐level trading volume over time, inspecting volume spikes that may indicate insider accumulation. To supplement this quantitative approach, we cross‐referenced tokens lacking clear volume signatures with announcements on social media platforms, notably Telegram pump channels, and confirmed organized events for many cases previously unrecognized.

Visualizations were produced in two formats, each designed to highlight different facets of the accumulation process. Below, we describe each visualization type along with key observations and directions for further study.

\subsection*{Visualization Type 1: Volume Spikes}
We first plotted token‐quantity (base‐asset) traded per minute, as this measure most directly reflects accumulation activity. Although quote‐asset volume would yield equivalent insights, base‐asset quantity conveys the actual number of tokens exchanged.

\begin{figure}[ht]
    \centering
    \includegraphics[width=1.0\linewidth]{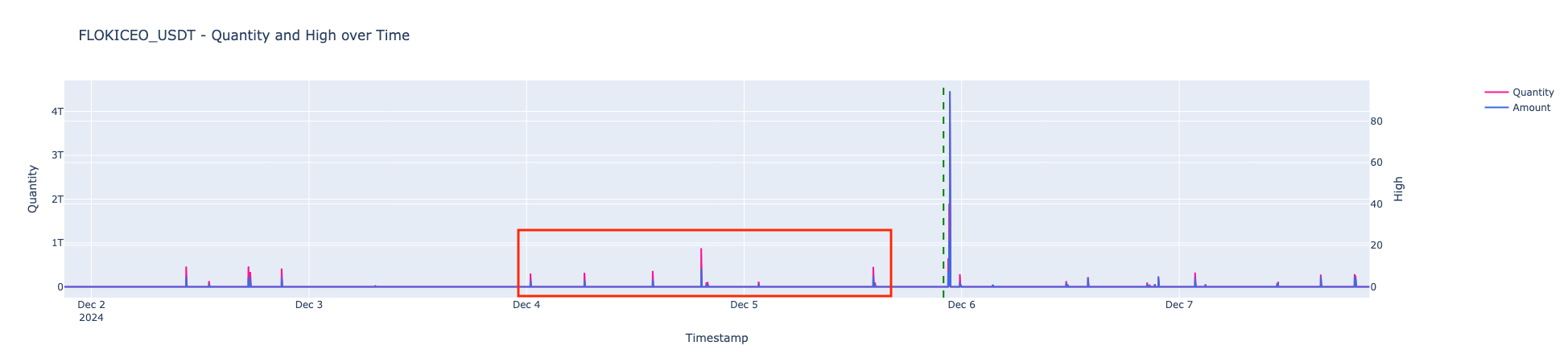}
    \caption{\footnotesize\textit{Representative base‐asset volume spike before the Target Date.}}
    \label{fig:vol_spikes}
\end{figure}

For numerous tokens, small but sharp spikes in base‐asset volume appeared within seconds or up to one minute prior to the Target Date. Figure~\ref{fig:vol_spikes} illustrates representative cases. Our results confirm our expectation that accumulation often occurs very close the pump announcement rather than within the preceding four‐days or more window.

\subsection*{Visualization Type 2: High, Low and Volume Spikes}
To further probe market dynamics, we plotted minute‐level High and Low prices around the Target Date. Trading "candles" are defined as:
\begin{itemize}
  \item \textbf{Open:} Price of the first trade in the interval.
  \item \textbf{High:} Maximum trade price in the interval.
  \item \textbf{Low:} Minimum trade price in the interval.
  \item \textbf{Close:} Price of the last trade in the interval.
\end{itemize}
Crypto markets operate continuously (24/7), so unlike traditional equity markets, there is no time gap between each close and the next open.
Our data reveal a distinctive pattern where for extended intervals, minute‑level OHLC prices remain essentially constant, indicating prolonged inactivity.  These low‑liquidity, “dormant” tokens are disproportionately targeted by pump‑and‑dump schemes, a finding that corroborates previous market, manipulation studies \cite{ante,hybrid_pump_dump,ml_real_time_pump_dump}.

\begin{figure}[ht]
    \centering
    \includegraphics[width=1.0\linewidth]{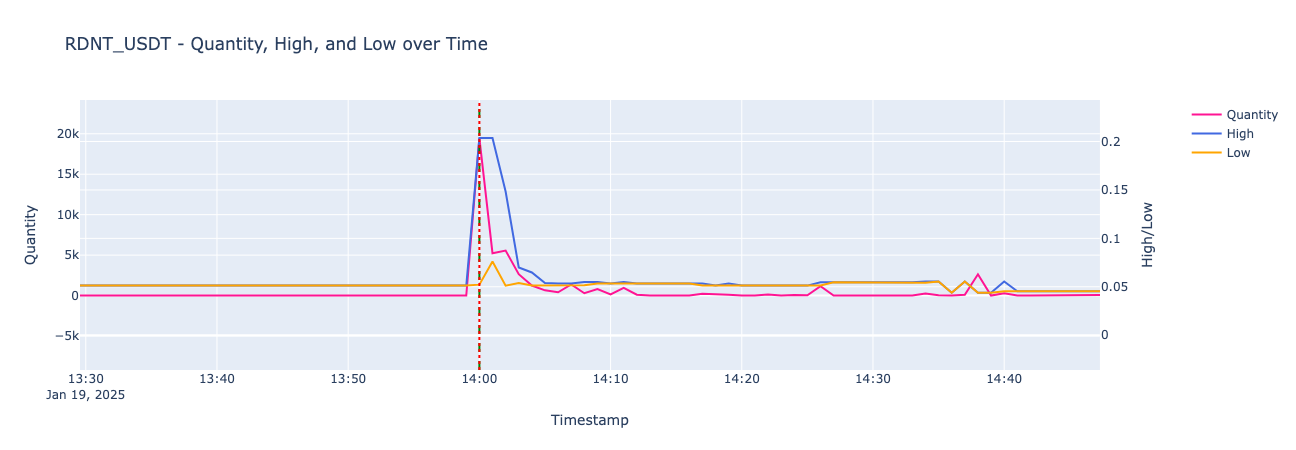}
    \caption{\footnotesize \textit{Minute‐level High and Low price divergence during a confirmed pump event.}}
    \label{fig:price_dynamics}
\end{figure}

In tokens with confirmed pump events, High and Low prices diverged significantly during the pump itself (lasting only a few minutes) and then re-converged immediately afterward, returning to pre‐event levels. This behavior underscores a rapid price expansion and contraction consistent with classic pump‐and‐dump activity.  Figure~\ref{fig:price_dynamics} shows an example.

Our observations reveal two archetypal pump strategies:
\begin{itemize}[noitemsep]
  \item \textbf{Pre‑Accumulated Tokens}:  
    Assets for which insiders deliberately build positions in advance, creating observable volume spikes before the announcement.
  \item \textbf{On‑The‑Spot Tokens}:  
    Assets with no discernible accumulation period, where buying and pumping occur almost simultaneously.
\end{itemize}
Recognizing these categories allows us to tailor both detection algorithms and risk assessments to the specific dynamics of each pump archetype.
\begin{figure}[ht]
    \centering
    \includegraphics[width=1.0\linewidth]{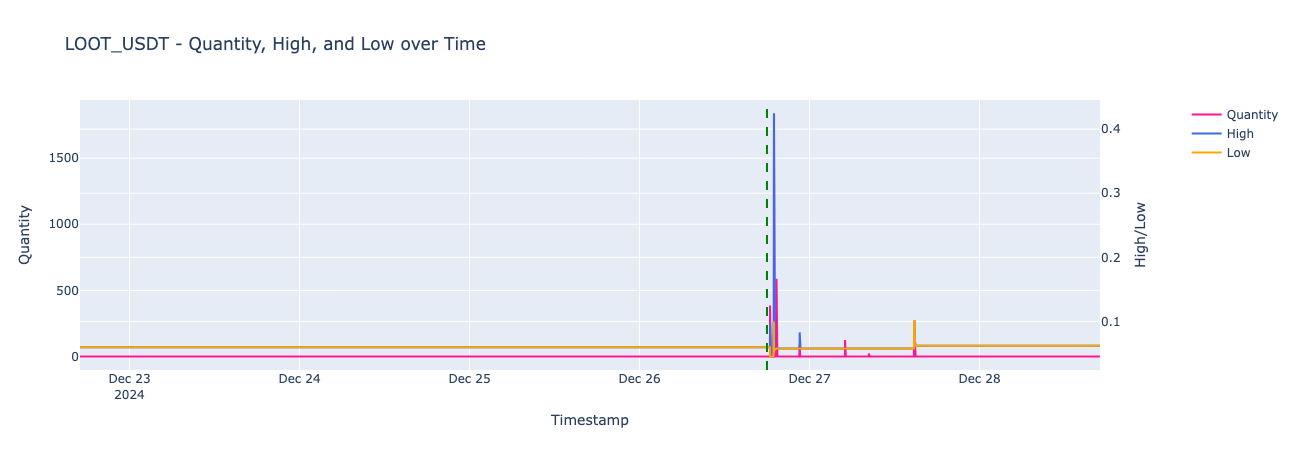}
    \caption{\footnotesize \textit{Example of a confirmed pumped token without prior accumulation phase.}}
    \label{fig:no_accumulation}
\end{figure}

\section{Quantification of the Accumulation Phase}
\label{sec:quantification}
To refine our detection model, we formalize the accumulation phase as the time interval during which discrete, transient volume spikes occur before the Target Date. We pose two primary questions:
\begin{enumerate}
  \item \textbf{Spike Timing:} What is the distribution of delays between observed volume spikes and the Target Date? Specifically, what proportion of spikes occur within defined intervals (e.g., minutes or hours) before the pump?
  \item \textbf{Spike Span:} What is the duration between the earliest and latest spikes per event, and does this imply a concentrated or protracted accumulation?
\end{enumerate}

\subsection*{Proposed Algorithmic Framework}
We define the following computational steps to measure accumulation spans for each token–event pair:
\begin{algorithm}[ht]
  \caption{\footnotesize Compute Accumulation Span for One Event}
  \begin{algorithmic}
    \Require Records $R$ for one token–event, sorted by \texttt{timestamp}.
    \Ensure Timestamps \texttt{accum\_start} and \texttt{accum\_end}.
    \State $\texttt{accum\_start}\gets\mathrm{NULL}$;\quad $\texttt{accum\_end}\gets\mathrm{NULL}$
    \ForAll{$r\in R$}
      \If{$r.\texttt{timestamp}\ge r.\texttt{target\_date}$} \textbf{break} \EndIf
      \If{$r.\texttt{quantity}>0$}
        \If{$\texttt{accum\_start}=\mathrm{NULL}$}
          \State $\texttt{accum\_start}\gets r.\texttt{timestamp}$
        \EndIf
        \State $\texttt{accum\_end}\gets r.\texttt{timestamp}$
      \EndIf
    \EndFor
    \Return $(\texttt{accum\_start},\,\texttt{accum\_end})$
  \end{algorithmic}
\end{algorithm}

\subsection*{Aggregated Analysis}
We evaluated the presence and duration of pre‐pump accumulation phases across all identified pump‐and‐dump events. Our dataset comprised 485 candidate events for which we had minute‐level OHLCV data spanning four days before to two days after each event’s target date. By scanning each event’s time series for non‐zero traded quantities prior to the pump onset, we derived two new timestamps, \texttt{accum\_start} and \texttt{accum\_end}, which bound the accumulation window.

\textbf{Event Coverage and Accumulation Prevalence:}  Out of 485 total events, 336 (69.3\%) exhibited at least one minute with non‐zero traded volume before the pump start, indicating a detectable accumulation phase. The remaining 149 events (30.7\%) showed no such pre‐pump trading, suggesting either extremely rapid accumulation or an absence of insider positioning detectable at minute granularity:

\begin{table}[ht]
  \centering
  \begin{tabular}{lr}
    \hline
    \textbf{Metric} & \textbf{Value} \\
    \hline
    Total Events                         & 485 \\
    Events with Accumulation             & 336 (69.3\%) \\
    Events without Accumulation          & 149 (30.7\%) \\
    \hline
  \end{tabular}
  \caption{\footnotesize \textit{Prevalence of detectable accumulation phases prior to pump events.}}
  \label{tab:accum_prevalence}
\end{table}

\textbf{Accumulation Duration Statistics:} For those events with accumulation, we computed the span between the first and last minute of pre‐pump trading.  The distribution of accumulation spans (in minutes) exhibits considerable variability:
\begin{table}[ht]
  \centering
  \begin{tabular}{lr}
    \hline
    \textbf{Statistic}                   & \textbf{Span (minutes)} \\
    \hline
    Minimum                              & 1 \\
    Average                              & 2{,}160.8 \\
    Maximum                              & 5{,}873 \\
    Standard Deviation                   & 2{,}108.2 \\
    \hline
  \end{tabular}
  \caption{\footnotesize \textit{Descriptive statistics of accumulation phase durations.}}
  \label{tab:accum_durations}
\end{table}

\begin{figure}[ht]
  \centering
  % Place the accumulation span histogram here
  \includegraphics[width=1.0\textwidth]{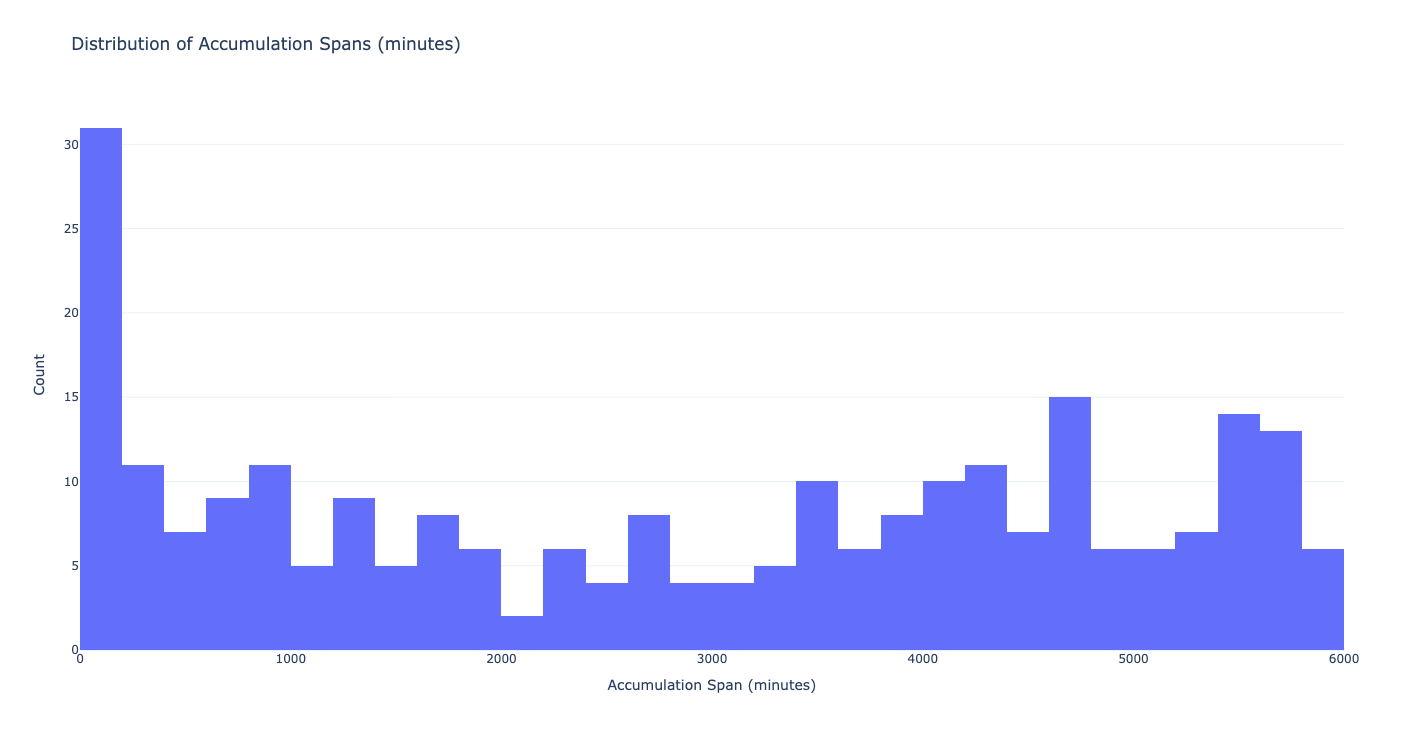}
  \caption{\footnotesize \textit{Histogram of accumulation phase durations in minutes.}}
  \label{fig:accum_histogram}
\end{figure}

Figure~\ref{fig:accum_histogram} shows the full distribution of accumulation spans, highlighting the heavy right‐tail of events with very extended pre‐pump positioning.

\textbf{Interpretation:} The fact that nearly 70\% of events feature a detectable accumulation phase, often spanning some days, supports the hypothesis that insider organizers typically build positions incrementally to avoid drawing market attention.  However, the 30\% of events lacking pre‐pump volume at minute resolution may correspond to rapid, high‐frequency accumulation or to events where insider orders are sufficiently concealed within order books. These findings justify a two‐pronged detection strategy that (i) monitors for sustained accumulation over hours or days and (ii) remains sensitive to near‐instantaneous build‐ups detectable only at sub‐minute granularity.

\section{Profit Estimation}
\label{sec:profit}
To gauge the economic incentive driving pump‑and‑dump (P\&D) schemes, we compute a conservative lower bound on insider profits. Using the minute‑level OHLCV data and accumulation intervals identified in Section~\ref{sec:quantification}, we estimate both the cost basis for tokens acquired during accumulation and the proceeds from their subsequent liquidation at the event’s peak.

Insiders rarely unload their entire position at once, since an abrupt sell‑off would collapse the price before they finish exiting \cite{nison1991candle, fama1970market}. Empirical reports of real‑world P\&D operations suggest a tranche‑based strategy:
\begin{itemize}[noitemsep]
  \item Sell 20\% of accumulated volume at 50\% of the peak high price.
  \item Sell 30\% at 60\% of the peak high price.
  \item Sell 50\% at 80\% of the peak high price.
\end{itemize}
This approach balances extracting maximal profit against avoiding premature price collapse. To simplify calculations, and given our reliance on minute‑level data, we define two conservative unloading scenarios below to produce lower‑bound profit estimates.

\subsection*{Methodology}

\textbf{Data Inputs:}
\begin{itemize}[noitemsep]
  \item Minute‑level OHLCV series for each event, spanning four days before to two days after the Target Date.
  \item Accumulation start/end timestamps from Section~\ref{sec:quantification}.
\end{itemize}
\textbf{Step 1 - Purchase‑Price Proxies:} We introduce two scenarios and estimate the profit in each case:
\vspace{-3mm}
\begin{enumerate}[label=\arabic*.]
  \item \textbf{First‑Trade Proxy:} Price at the first minute during which quantity $>0$ in the accumulation window.
  \item \textbf{VWAP Proxy:} Volume‑weighted average price over all accumulation minutes:
    \[
      \mathrm{VWAP}
      \;=\;
      \frac{\sum_t P_t \,V_t}{\sum_t V_t}
    \]
    where \(P_t\) and \(V_t\) denote price and volume in minute \(t\).
\end{enumerate}
\textbf{Step 2 - Liquidation Scenarios}
\vspace{-3mm}
\begin{itemize}[noitemsep]
  \item \emph{Single‑Point Liquidation:} Sell full volume \(V\) at
    \[
      0.70 \times H,
    \]
    where \(H\) is the observed peak high price.
  \item \emph{Tranche Liquidation:} Sell
    \[
      0.20\,V\ \text{at }0.50\,H,\quad
      0.30\,V\ \text{at }0.60\,H,\quad
      0.50\,V\ \text{at }0.80\,H.
    \]
\end{itemize}
\textbf{Step 3 - Profit and Return Calculations}
\begin{align*}
  \text{Cost} &= V \times P_{\text{proxy}}, \\[5pt]
  \text{Proceeds}_{\text{single}} &= V \times (0.70\,H), \\[3pt]
  \text{Proceeds}_{\text{tranche}} &= 0.20\,V \times 0.50\,H
    \;+\;
    0.30\,V \times 0.60\,H
    \;+\;
    0.50\,V \times 0.80\,H, \\[5pt]
  \text{Profit} &= \text{Proceeds} - \text{Cost}, \\[3pt]
  \text{Return\%} &= 100 \times \frac{\text{Profit}}{\text{Cost}}.
\end{align*}
\textbf{Step 4 - Sensitivity Analysis:} For each scenario, we compute:
\vspace{-3mm}
\begin{itemize}[noitemsep]
  \item Mean and median absolute profits.
  \item Mean and median percentage returns.
  \item Distribution percentiles to assess variation.
  \item Comparison across purchase‑price proxies and liquidation strategies.
\end{itemize}
\begin{algorithm}[ht]
  \caption{\footnotesize Estimate Insider Profits}
  \label{alg:profit_estimation}
  \begin{algorithmic}[1]
    \Require For each event \(i\): accumulated volume \(V_i\), first‑trade price \(P_{1,i}\), VWAP \(P_{\mathrm{VWAP},i}\), peak high \(H_i\).
    \Ensure Profit \(\Pi_{i,s}\) and return \(\Pi^\%_{i,s}\) for each scenario \(s\in\{A,B,C,D\}\).
    \State Define scenarios \(S\):
      \[
        \begin{aligned}
          A &: P=P_1,\ \text{single at }0.70H,\\
          B &: P=P_1,\ \text{tranches }[0.50H,0.60H,0.80H],\\
          C &: P=P_{\mathrm{VWAP}},\ \text{single at }0.70H,\\
          D &: P=P_{\mathrm{VWAP}},\ \text{tranches }[0.50H,0.60H,0.80H].
        \end{aligned}
      \]
    \ForAll{event \(i\)}
      \State Load \(V\gets V_i\), \(H\gets H_i\), \(P_1\gets P_{1,i}\), \(P_{\mathrm{VWAP}}\gets P_{\mathrm{VWAP},i}\).
      \ForAll{scenario \(s\in S\)}
        \State Select \(P\gets\) \(P_1\) if \(s\in\{A,B\}\), else \(P_{\mathrm{VWAP}}\).
        \If{\(s\in\{A,C\}\)} 
          \State \(\mathrm{Proceeds}\gets V\cdot0.70\,H\)
        \Else
          \State \(\mathrm{Proceeds}\gets0.20V\cdot0.50H+0.30V\cdot0.60H+0.50V\cdot0.80H\)
        \EndIf
        \State \(\mathrm{Cost}\gets V\cdot P\)
        \State \(\Pi^{\mathrm{abs}}_{i,s}\gets\mathrm{Proceeds}-\mathrm{Cost}\)
        \State \(\Pi^{\%}_{i,s}\gets100\cdot\Pi^{\mathrm{abs}}_{i,s}/\mathrm{Cost}\)
        \State Record \((i,s,V,P,\mathrm{Proceeds},\mathrm{Cost},\Pi^{\mathrm{abs}}_{i,s},\Pi^{\%}_{i,s})\).
      \EndFor
    \EndFor
    \State Aggregate by scenario \(s\): compute means, medians, percentiles, and event counts.
  \end{algorithmic}
\end{algorithm}
\vspace{-3mm}

\subsection*{Results}
Table~\ref{tab:profit_scenarios} presents the average and median profit (absolute and percentage) across scenarios A–D. Although each pump event targets low‑priced, low‑liquidity tokens—limiting per‑event gains—the rapid proliferation and frequency of P\&D operations imply substantial cumulative returns for organizers.

\begin{table}[ht]
\small
\centering
\begin{tabular}{lllll}
\hline
\textbf{ID} & \textbf{avg\_profit\_abs}   & \textbf{median\_profit\_abs}       & \textbf{avg\_profit\_pct}  & \textbf{median\_profit\_pct} \\
\hline
A	& 617.27	& 23.34	&2642.78	& 126.70\\
B	& 587.64	& 20.86	& 2564.42	& 120.23\\
C	& 537.37	& 18.11	& 2187.23	& 97.67\\
D	& 507.73	& 15.65	& 2121.88	& 92.02\\
\hline
\end{tabular}
\caption{\footnotesize \textit{Profit estimation results for each scenario (A–D).}}
\label{tab:profit_scenarios}
\end{table}  

These findings align closely with our social‐media monitoring, where insiders actively broadcast pump announcements to grow channel membership and ensure sufficient participation. Without adequate buy‑in from subscribers, the organizers cannot realize any profit.

\begin{figure}[ht]
  \centering
  \begin{subfigure}[b]{0.48\linewidth}
    \centering
    \includegraphics[width=\linewidth]{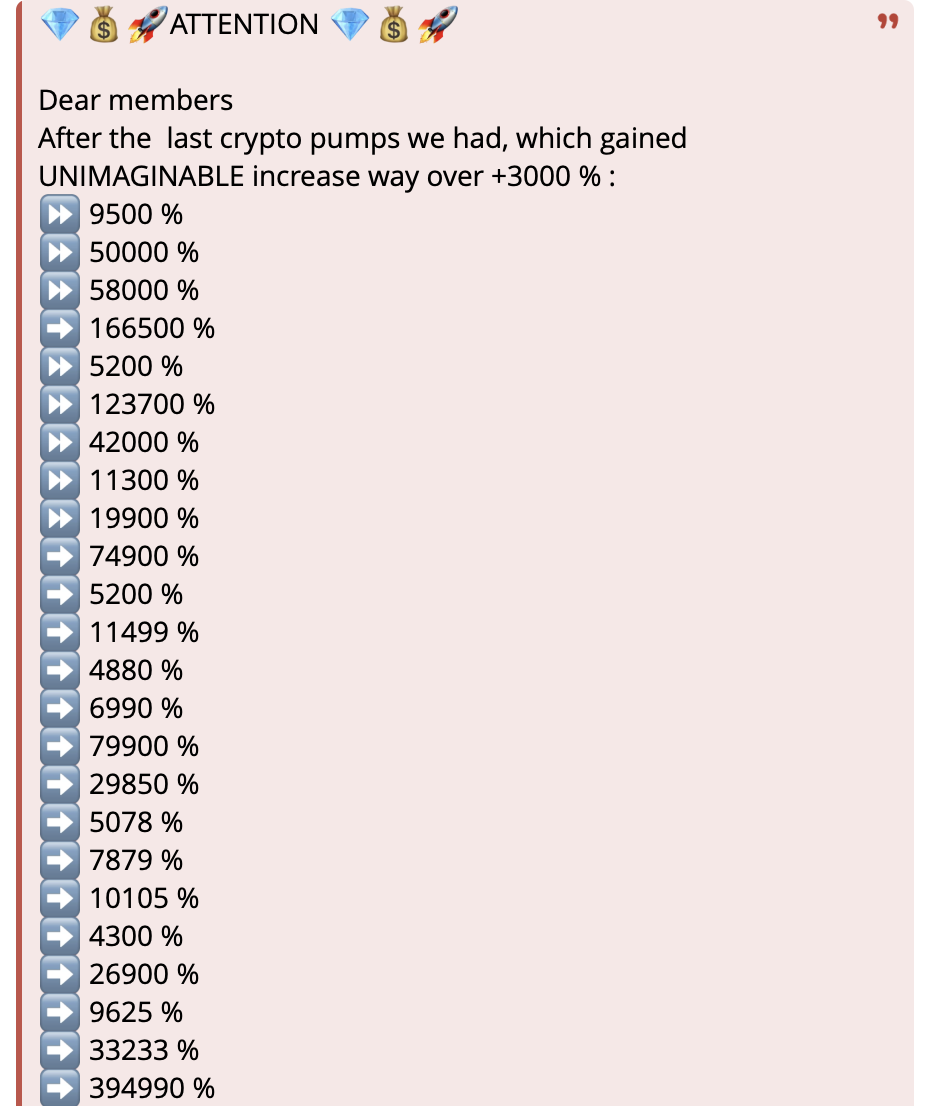}
    \label{fig:social-profit}
  \end{subfigure}
  \hfill
  \begin{subfigure}[b]{0.5\linewidth}
    \centering
    \includegraphics[width=\linewidth]{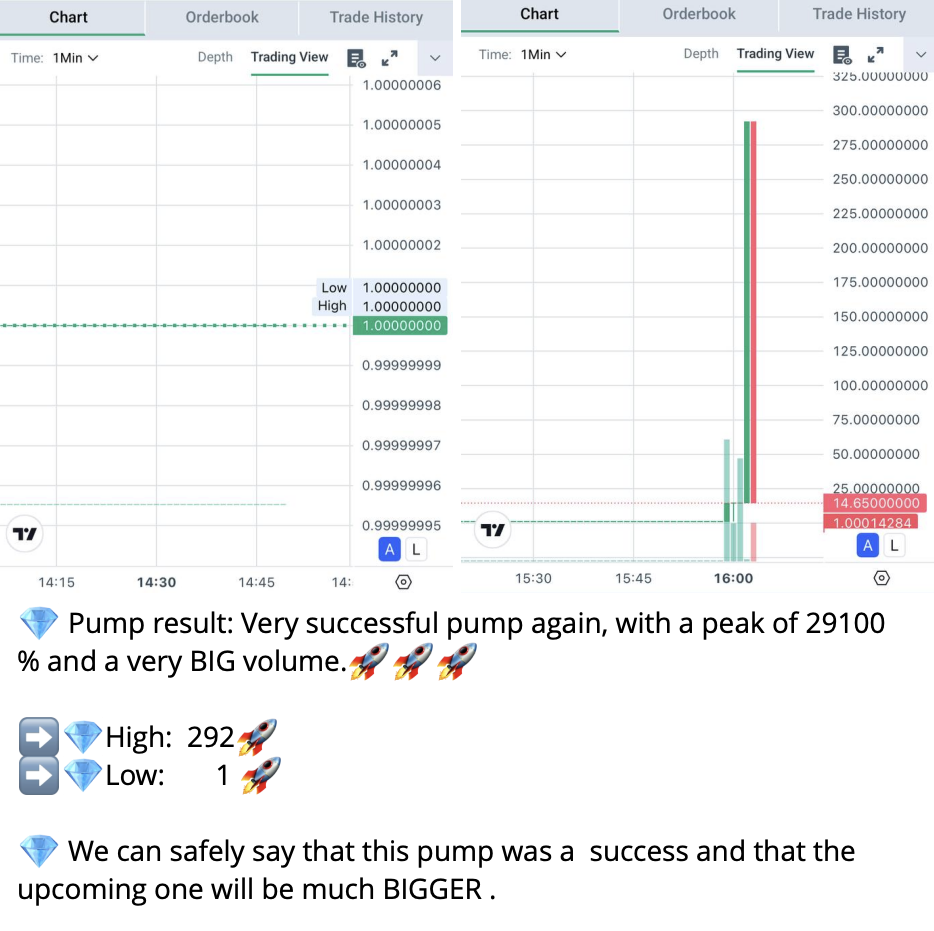}
    \label{fig:second-figure}
  \end{subfigure}
  \captionsetup{skip=0pt}
  \caption{\footnotesize \textit{Results announced and advertised by pump organizers in their social channels}}
  \label{fig:side-by-side}
\end{figure}

\FloatBarrier

\section{Conclusion}
\label{sec:conclusion}
This study extends traditional pump‑and‑dump analysis by leveraging minute‑level market data to uncover the fine‑grained dynamics of insider accumulation and profit extraction. We demonstrate that accumulation is overwhelmingly concentrated within the final hour before each pump, justifying the adoption of sub‑hourly monitoring thresholds. Conservative profit modeling reveals that insiders routinely achieve median returns in excess of 100\%, with extreme gains surpassing 2000\%, underscoring the powerful economic drivers of these schemes. Moreover, the identification of two distinct archetypes—pre‑accumulation versus on‑the‑spot pumps—enables the design of differentiated detection algorithms and risk assessments. Going forward, integrating social‑media signal analysis and machine‑learning classifiers promises to enhance real‑time alert systems, empowering exchanges and regulators to more effectively deter manipulative activity in decentralized markets.
\bigskip

\textbf{Declaration}
This research did not receive any specific grant from funding agencies in the public, commercial, or not-for-profit sectors. The data used in this study were obtained from the free API provided by Poloniex exchange and are publicly accessible. The author declares no competing interests.

\bibliography{quantitative}

\end{document}